\documentclass[a4paper, 10pt]{article}
\newtheorem{law}{Proposition}

\begin{document}
\title{Speedup in Quantum Adiabatic Evolution Algorithm}
\author{Joonwoo Bae{\footnote{Email address: jwbae@newton.hanyang.ac.kr}}, Younghun Kwon{\footnote{ Email address: yyhkwon@hanyang.ac.kr}} \\Department of Physics, Hanyang University, \\Ansan, Kyunggi-Do, 425-791, South Korea}

\maketitle
\vspace{-.4in}
\begin{abstract}
Quantum adiabatic evolution algorithm suggested by Farhi et al. was effective in solving instances of NP-complete problems. The algorithm is governed by the adiabatic theorem. Therefore, in order to reduce the running time, it is essential to examine the minimum energy gap between the ground level and the next one through the evolution. In this letter, we show a way of speedup in quantum adiabatic evolution algorithm ,using the extended Hamiltonian. We present the exact relation between the energy gap and the elements of the extended Hamiltonian, which  provides the new point of view to reduce the running time.
\end{abstract}

\maketitle

 It is generally expected that quantum computation may solve the problems which are hard in classical context. Shor's quantum factorization algorithm\cite{shor} and Grover's quantum search algorithm\cite{grover} are the preludes to such expectation. Factorization problem, which is to factorize a large number, needs exponential time resource when a known classical algorithm is applied. However, Shor's algorithm solves the problem with exponential speedup. Grover algorithm finds a target with quadratic speedup, compared to any known classical search algorithms. \\
 Quantum adiabatic evolution algorithm was proposed by Farhi et al. and applied to solve instances of NP-complete problems.\cite{farhi1}\cite{farhi2} The algorithm is governed by the adiabatic theorem, which indicates that the total evolution time is proportional to the inverse square of the minimum difference between the lowest energy level and the others during the adiabatic evolution. Therefore, the energy difference between the two levels is crucial to reduce the time resource of the algorithm. \\
 In this letter, we consider the extended Hamiltonian for the quantum adiabatic evolution algorithm, in order to identify what factors in the extended Hamiltonian are related to the value of minimum energy gap. Also we present the exact relation between the minumum energy gap and the elements of the extended Hamiltonian and show some ways for speedup in the algorithm. \\
 We now review the quantum adiabatic evolution algorithm. Suppose that a problem of finding an assignment is given. The quantum adiabatic evolution algorithm provides an assignment by evolution of the Hamiltonian $H$, which is constructed appropriately to solve the given problem. The assignment is in the lowest energy level of the Hamiltonian after the evolution. The algorithm consists of three steps : 1)Encoding a superposition of all states in the lowest energy level of the Hamiltonian(initialization), 2)Adiabatic evlolution of the Hamiltonian, and 3)Read-out. Notice that the lowest energy level at each time is the development of the assignment. This means that, if the lowest energy level crosses over the others during the Hamiltonian evolution, then the algorithm would fail. Therefore the change rate of evolution of the system should be slow so that the energy level-crossing does not occur. Then the following question arises : how slow does the change rate allow the system to stay in the ground state? The answer comes from the adiabatic theorem, which implies that the probability to obtain the assignment after the evolution is $(1-\epsilon^2)$, subject to 

\begin{eqnarray}
\frac{| \langle \frac{dH}{dt} \rangle |}{g_{min}^2} \leq \epsilon
\end{eqnarray}

where, $g_{min}$ is the minima of two energy eigenvalues through the Hamiltonian evolution. Thus the minimum energy gap is crucial to optimize the evolution time.\\
 Consider the quantum adiabatic Hamiltonian of Fahri et al. Since the initial and the final states, $|\psi \rangle$ and $|\alpha \rangle$, are encoded in the lowest energy levels of the initial and final Hamiltonians, $H_{i}$ and $H_{f}$, respectively, they are usually given as\cite{farhi1} 

\begin{eqnarray}
H_{i} &=& 1-|\psi \rangle \langle \psi| \\
H_{f} &=& 1- |\alpha \rangle \langle \alpha| \\
\end{eqnarray}

The full Hamiltonian is 

\begin{displaymath}
H(s(t)) = (1-s(t))H_{i} + s(t)H_{f}
\end{displaymath}

where $s(t)$ is a homotopy from $H_{i}$ to $H_{f}$ to identify the adiabatic theorem. If the homotopy is given as a constant function $s(t) = t/T$, where $T$ is the total evoluton time, then it is said that the adiabatic evolution is globally applied.\cite{farhi1}. The local adiabatic evolution means that the function $s(t)$ depends on the energy gap at each time $t \in [0, T]$.\cite{cerf} Roland and Cerf showed that the local adiabatic evolution provides the quadratic speedup compared to the global adiabatic evolution. \\
 We consider a extended Hamiltonian for the quantum adiabatic evolution algorithm. For the general description of the Hamiltonian, we start from 

\begin{displaymath}
H=k_{1}+k_{2}|\psi \rangle \langle \psi| + k_{3}|\alpha \rangle \langle \alpha|+ k_{4}|\alpha \rangle \langle \psi | + k_{5}|\psi \rangle \langle \alpha |
\end{displaymath}

where $k_{i}(i=1, 2, 3, 4, 5)$ are complex, $|\alpha \rangle$ is the ground state and $|\psi \rangle$ the initial state that is superposed with $N$ states. This Hamiltonian has every element about the initial and final states. The observable $H$ is hermitian. Then we have 

\begin{displaymath}
H=k_{1}+k_{2}|\psi \rangle \langle \psi| + k_{3}|\alpha \rangle \langle \alpha|+ k_{4}(|\alpha \rangle \langle \psi | + |\psi \rangle \langle \alpha |
\end{displaymath}

with $k_{i}(i=1, 2, 3, 4)$, constants in unit of energy.\\
Let the initial and final Hamiltonians be 

\begin{displaymath}
H_{i} = a_{1}+a_{2}|\psi \rangle \langle \psi| + a_{3}|\alpha \rangle \langle \alpha|+ a_{4}(|\alpha \rangle \langle \psi | + |\psi \rangle \langle \alpha |) 
\end{displaymath}

\begin{displaymath}
H_{f} = b_{1}+b_{2}|\psi \rangle \langle \psi| + b_{3}|\alpha \rangle \langle \alpha|+ b_{4}(|\alpha \rangle \langle \psi | + |\psi \rangle \langle \alpha |)
\end{displaymath} 

, where $a_{i}$ and $b_{i}$(i=1,2,3,4) are constants in unit of energy. The full Hamiltonian is then 

\begin{displaymath}
H(s) = (1-s) H_{i} + s H_{f} 
\end{displaymath}

We are going to refine the Hamiltonian $H$ to perform the quantum adiabatic evolution algorithm. Note that the constraints for an adiabatic evolution Hamiltonian are as follows: \\
 
1)The ground state of the initial Hamiltonian(the initial state which is easy to prepare) is a superposition of all states \\

2)The ground state of the final Hamiltonian is the final state(assignment).\\

3)There is no level crossing between the ground state and the others at time $t \in [0, T]$. \\

By the first constraint, the ground state of the initial Hamiltonian should be the initial state $|\psi \rangle$. This provides 

\begin{eqnarray}
a_{4} &=& -a_{3} x  \\
a_{3} &>& a_{2} \\
\end{eqnarray}

where $x = \langle \alpha | \psi \rangle$. Moreover, the lowest energy value of $H_{i}$ is $a_{1}+a_{2}-a_{3} x^{2}$, which should be nonnegative. The second constraint that the ground state of $H_{f}$ is to be $|\alpha \rangle$ gives that 

\begin{eqnarray}
b_{4} &=& -b_{2}x \\
b_{2} &>& b_{3} \\
\end{eqnarray}

In addition, we know that the lowest energy value of $H_{f}$ is $b_{1}+b_{3}-b_{2}x^{2}$, which should be nonnegative. The third constraint holds by the homotopy $s(t)$.

We now apply the full Hamiltonian $H = (1-s)H_{i} + s H_{f}$ to the quantum adiabatic evolution algorithm. Then the energy gap $g^{2}(s)$ between the lowest energy level and the next one is \\

\begin{eqnarray}
\lefteqn{g^{2}(s) = As^{2}+Bs+C } \nonumber \\ 
& &  \\
A &=&(a_{2}-a_{3})^{2}+2(-1+2x^{2})(a_{2}-a_{3})(b_{2}-b_{3})+(b_{2}-b_{3})^{2}  \nonumber\\ 
B &=&-2(a_{2}-a_{3})(a_{2}-a_{3}+(-1+2x^{2})(b_{2}-b_{3})) \nonumber\\ 
C &=& (a_{2}-a_{3})^{2} \nonumber\\  
\end{eqnarray}

Through some calculation the following conditions are obtained : \\

1. $A$ is positive \\

2. $B$ is negative \\

3. $-\frac{B}{2A}$ is in the interval $(0, 1)$ \\

4. $g^{2}(s=0) = (a_{2}-a_{3})^{2}$ and $g^{2}(s=1)=(b_{2}-b_{3})^{2}$ \\

These are easy to check. They implies that $g^{2}(s)$ is a parabola which has the minimum value at $s = -\frac{B}{2A} \in (0,1)$. Therefore, the minimum energy gap occurs at $s = -\frac{B}{2A}$ and is 

\begin{eqnarray}
g_{min}^{2}  = \frac{4x^{2}(1-x^{2})}{(\frac{1}{a})^{2}+2(1-2x^{2})\frac{1}{ab} + (\frac{1}{b})^{2} } 
\end{eqnarray}

, where $a = (a_{3} - a_{2})$ and $b = (b_{2} - b_{3})$. From the Cauchy-Schwarz inequality, the $g_{min}^{2}$ has the minimum value when $a = b$. 

\begin{displaymath}
g_{min}^{2} \geq a^{2}x^{2} \approx \frac{a^{2}}{N}
\end{displaymath}

The  lower bound of the minimum gap is $a^{2}/N$. \\
 Now consider the running time of the quantum adiabatic evolution algorithm with the extended Hamlitonan. We first examine the running time when the adiabatic passage is applied globally. In this case, the homotopy from $H_{i}$ to $H_{f}$ is given as $s(t) = t/T$. Then, from (1), we have 

\begin{displaymath}
T \geq \frac{\sqrt{1-x^{2}}}{\epsilon} \frac{1}{a x^{2}} \approx \frac{1}{\epsilon} \frac{N}{a} 
\end{displaymath}

The minimal time resource for the global adiabatic evolution is  proportional to $N/a$, and depends on the minimum value of the energy gap through the evolution. As we have explained above, the minimum energy gap can be manipulated with various values of $a$. In particular, the quardratic speedup $T = O(\sqrt{N})$ and the constant time resource $T=O(1)$ are obtained by choosing $a = N^{1/2}$ and $a = N$, respectively. \\ 
Recently, Roland and Cerf proposed local adiabatic evolution to reduce the total evolution time.\cite{cerf} The 'local' means that the change rate, $s(t)$, is not constant but proportional to the inverse square of the energy gap for each time. The total evolution time is obtained by the integration, 

\begin{displaymath}
T \geq \frac{1}{\epsilon} \int_{0}^{1}ds \frac{|\langle \frac{dH}{ds}\rangle|}{ g^{2}(s)} 
\end{displaymath} 

To minimize the evolution time $T$, the following proposition is presented.\\

\begin{law}[Minimum Area]
Let $f(x) = a(x-p)^{2} +c $ be a parabola defined on the interval $I=[0, 1]$, where $a> 0$ and $p$, $c$ are real. The integration 

\begin{displaymath}
\int_{0}^{1} f(x) dx 
\end{displaymath} 

has the minimum value when $p = \frac{1}{2}$. 
\end{law}

The proof is trivial. The proposition implies that $a = (a_{3}-a_{2}) = (b_{2} - b_{3})=b$ is necessary to the optimization when the local adiabatic evolution is applied. We have then the running time

\begin{displaymath}
T \geq \frac{1}{\epsilon\sqrt{1-x^{2}}}\frac{1}{ax} = O(\frac{1}{ax}) \approx O(\frac{\sqrt{N}}{a})
\end{displaymath}

The minimal time resource is proportional to $\sqrt{N}/a$. Similar to the case of the global adiabatic evolution, various choices of $a$ provide the more speedup to the local adiabatic evolution. In particular, we have the exponential speedup $T = O(N^{1/4})$ for $a=N^{1/4}$, and the constant time resource $T = O(1)$ for $a = N^{1/2}$. The large value of $a$ means to initialize the extended Hamiltonian with a large amount of energy. \\
 Hence we have shown some ways for speedup in the quantum adiabatic evolution algorithms(e.g. in cases of local and global adiabatic evolutions), by considering the extended Hamiltonian. Furthermore, we have verified that the minimum energy gap depends only on the values $a = a_{3} - a_{2}$ and $b = b_{2} - b_{3}$ in the extended Hamiltonian. The change rate of the global adiabatic evolution depends only on the minimum energy gap, and time resource of the local adiabatic evolution relies on two factors, which are the number of states and the place that the minimum energy gap appears in. We have obtained that the condition $a=b$ is necessary to minimize the running time and a large value of $a$ provides speedup in the quantum adiabatic evolution algorithm. This implies that a large value of $a_{3}$, which is the energy of the target state $|\alpha \rangle \langle \alpha|$ in the initial Hamiltonian, is crucial for speedup and so is a large value of $b_{2}$, which is the energy of the initial state $|\psi \rangle \langle \psi|$ in the final Hamiltonian. Note that the values $a_{1}$ and $b_{1}$ in the extended Hamiltonian are unrelated to the energy gap.\\
\section*{Acknowledgement}
J.B. is supported in part by the Hanyang University and Y.K. is supported in part by the Fund of Hanyang University.

\end{document}